\documentclass[aps,prx,onecolumn,showpacs,superscriptaddress]{revtex4}

\usepackage{epsfig}
\usepackage{graphicx}
\usepackage{enumitem}

\begin{document}

\title{Commentary: Intentional Observer Effects on
Quantum Randomness: A Bayesian Analysis Reveals Evidence Against
Micro-Psychokinesis}

\author{H. Grote}
\email{groteh@cardiff.ac.uk}
\affiliation{Cardiff University, School of Physics and Astronomy\\
The Parade, Cardiff, CF24 3AA, United Kingdom}

\raggedbottom

\date{\today}

\begin{abstract}
The paper titled `Intentional Observer Effects on
Quantum Randomness: A Bayesian Analysis Reveals Evidence Against
Micro-Psychokinesis',
published in Frontiers of Psychology in March 2018 \cite{Maier2018},
reports on a mind-matter experiment with the main result of
strong evidence against Micro-Psychokinesis.
Despite this conclusion, the authors
interpret the observed pattern in their data
as possible evidence for Micro-Psychokinesis, albeit of a different kind.
Suggesting a connection to some existing models, the authors put forward the hypothesis 
that a higher frequency of slow data variations can be observed in their experiment data
than in a set of control data. This commentary analyses this claim and concludes that
the variation in the data motivating this hypothesis would show up just by chance
with a probability of p=0.328 under a null hypothesis.
Therefore, there is no evidence for the hypothesis of faster data variations,
and thus for this kind of suggested Micro-Psychokinesis in this experiment.
\end{abstract}
  

\maketitle

In the work reported in \cite{Maier2018} the authors conduct an experiment with high statistical
power, testing 12.571 subjects in a Micro-PK task. The authors find strong evidence against
the existence of Micro-PK in their main analysis, testing the aggregate sum of their data
against the expectation value. Setting aside the question whether any study of this kind can be 'decisive' in this field, the reported work is impressive in its scope.

However, despite this negative finding the authors report an additional, post-hoc,
investigation in which they propose that PK-Effects show up in fluctuations of their data:
\emph{Interestingly, there seems to be a pattern of repeated change}.
The authors connect this possible observation with theories of von Lucadou 
(e.g.\cite{Lucadou2006}) and others,
extended by their own thoughts about possible decline effects in Parapsychology experiments.
It seems to this author that there is a confusion here about decline of the primary
effect size and a decline as observed in a cumulative z-score representation of data,
as used by the authors.
A constant oscillation of an original effect size
always leads to a decline of oscillations in a cumulative z-score plot, as more data
goes into the z-score calculation. Likewise, this confusion is also evident in Figure 7 of \cite{Maier2018}. The cumulative z-scores in those figures have a constant oscillation
amplitude, which is only possible with an oscillating and exponentially growing effect size 
in the original data.

Setting this commentary aside, the authors state that:
\emph{We propose that the data presented in this study here also follow a similar systematic pattern of decline matching dampened harmonic oscillation function as suggested by Maier and Dechamps (in press).}
The authors then move to the hypothesis that their experimental data
does show faster variations than expected by chance, thus supporting the existence of a 
particular form of PK-Effect.
To support this hypothesis the authors generate one set of control data, comprising the same amount of data as the experimental data set (i.e. 12.571 'simulated' participants).
They state:
\emph{Comparing Human and Simulated Data
The human and the simulated data should - if the harmonic oscillation assumption is true - 
differ mainly in the frequency parameter $\omega$. Real effects should produce more pronounced oscillations than artificial data. To explore this, we compared
the 95\%-confidence intervals for both frequency scores and
found indeed that they did not overlap.}

This last finding, however, does not say much about the hypothesis of the authors.
Deriving any significance from this single observation is incorrect. The 95\%-confidence
intervals of the fits of the oscillating functions have nothing to do with the question
whether there is a predominant frequency (of whatever quantity) in one dataset versus another one.
To illustrate this point one can imagine a single data set which is mostly composed of two different 
frequencies of almost equal amplitude. A fitting algorithm may find both solutions to be
reasonable fits, within some error margin for each fit. However, the decision which of the two fits is actually 'better' can be a marginal one.
What is required in the case the authors want to assess, is not only the comparison to one 
control data set, but to an ensemble of \emph{many} control data sets.

The generation of a large amount of control data sets does not necessarily have to be performed using the original apparatus, which may be a too time-consuming enterprise.
Random control data can be generated with deterministic algorithms in cases the statistical parameters
of the resembled experiment are sufficiently simple and well known.
Certainly this is the case for the experiment here, where only the sum of 100 binary decicions
constitute one datapoint per participant.
Another way to generate large amounts of control data is the use of permutations of the
original experimental data or control data, obtained with the original apparatus.

For illustration, this author has generated 1000 data sets from random permutations
of the 12571 data points of the experimental data as reported in \cite{Maier2018}.
Using Mathematica for the fitting of an oscillatory decaying function (the same function as noted on page 7 in \cite{Maier2018}), for each data set the fit was initiated with the following start values:
a=1, $\beta = 0$, and p, m, and h unspecified. The start value for the frequency parameter $\omega$
was chosen at random for each fit on the interval $0.0005 < \omega < 0.005$,
which is a range of frequencies of interest to test the hypothesis of the authors.
The frequency value $\omega$ is the variable under test. 

\begin{figure*}[h!]
\includegraphics[width=11cm]{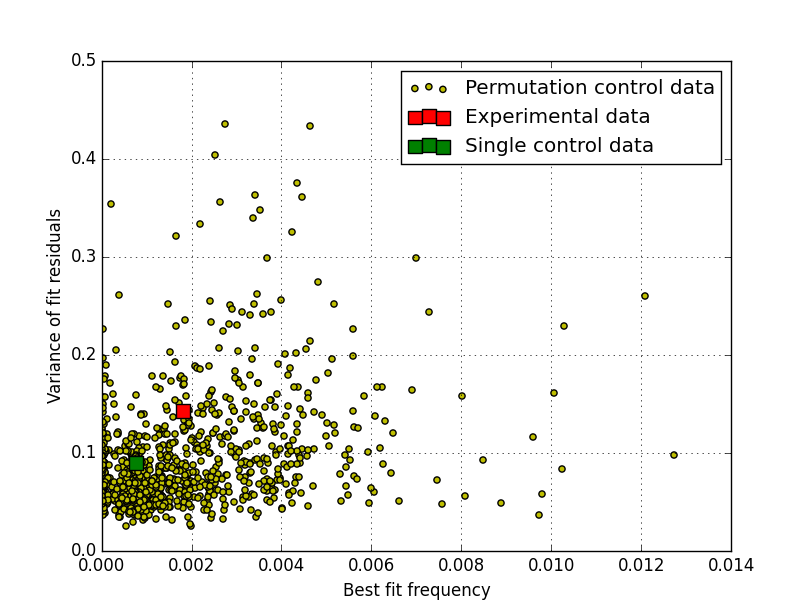}
\vspace{0mm}
\caption{Scatter plot of 1000 sets of permutated data that each have been fitted with
the (decaying) oscillation to determine the best fit frequency ($\omega$).
The main experimental result as presented in \cite{Maier2018} lies at $\omega = 0.0018$
with 386 results of the simulated data having larger $\omega$-values than that.
The majority of fits has better goodness-of-fit (smaller variance of the fit residuals)
than the experimental data as reported in \cite{Maier2018}. The (unpermuted) experimental data is shown as the red square and the single control data set of \cite{Maier2018} is
shown as a green square.
}
\label{mcdata}
\end{figure*}

The fit results are shown in Figure~\ref{mcdata} for the parameter $\omega$ in
form of a scatter plot. To assess the quality of each fit, the variance of the fit residulas has been evaluated for each fit and is plotted as associated parameter.
The main experimental result as presented in \cite{Maier2018} lies at $\omega = 0.0018$
with 386 results of the 1000 simulated data sets having larger $\omega$-values than that.
If one ignores fits with variance higher than 0.14 (the variance of the original data),
282 out of 858 results have higher frequencies than the experimental data.
In other words, the variation in the data motivating the hypothesis under test,
would show up just by chance with a probability of p=0.328 under a null hypothesis.
This result has qualitatively also been confirmed using the single control data set for the permutations, and also using a pseudo-random number generator, showing that this finding is robust with respect to the source of randomness.


This author concludes that there is no evidence for the hypothesis that the experimental
data reported in \cite{Maier2018} show faster oscillations than expected by chance.
Random permutations of the original data produce many data sets with even higher
dominant frequencies, as illustrated in this commentary.
The main point of this commentary is that a \emph{distribution} of many control data sets has to be used, in order to assess the statistical significance of the hypothesis under test by the authors.








\end{document}